\documentclass[12pt]{iopart}

\usepackage{iopams}  
\usepackage {cancel} 
\usepackage{cite} 
\usepackage{xcolor}

\begin{document}

\title[]{Isomorphism between the Bia{\l}ynicki-Birula and the Landau-Peierls Fock space quantization
of the electromagnetic field in position representation}

\author{M. Federico, H. R. Jauslin}

\address{Laboratoire Interdisciplinaire Carnot de Bourgogne, UMR 6303 CNRS\\- Universit\'e de Bourgogne Franche-Comté, BP47870, 21078 Dijon, France}
\ead{jauslin@u-bourgogne.fr}
\vspace{10pt}
\begin{indented}
\item[]\today
\end{indented}

\begin{abstract}
We first present a summary of the quantization of the electromagnetic field in position space representation, using two main approaches: the Landau-Peierls approach in the Coulomb gauge and the Bia{\l}ynicki-Birula approach, based on the Riemann-Silberstein vector. We describe both in a framework that starts with a classical Hamiltonian structure and builds the quantum model in a bosonic Fock space by a precisely defined principle of correspondence. We show that the two approches are completly equivalent. This is formulated by showing that there is a unitary map between the Fock spaces that makes them isomorphic. Since all the physically measurable quantities can be expressed in terms of scalar products, this implies that the two quantizations lead to exactly the same physical properties. We show furthemore that  the isomorphism is preserved in the time evolutions. To show the equivalence, we use the concepts of helicity and frequency operators. The combination of these two operators provides a formulation that allows one to make the link between these two methods of quantization in a precise way. We also show that the construction in the Bia{\l}ynicki-Birula quantization that avoids the presence of negative eigenvalues in the Hamiltonian, in analogy with the one for the Dirac equation for electrons and positrons, can be performed through an alternative choice of the canonical variables for Maxwell's equations.
\end{abstract}

%
\vspace{2pc}
\vspace{2cm}
\\
%
\maketitle
%
%

\section{Introduction}

In the past few years, many headways have been made in single-photon production, such as the designing of the pulse shapes \cite{Kuhn2002,McKeever2004,Keller2004}, which provides a powerful tool for the development of new quantum technologies. This feature has brought back the interest of the description of photons in the position space representation \cite{Sipe1995,Keller2000,Chan2002,Keller2005,Smith2007} and of photon localization \cite{Hawton2001,Saari2005,Gulla2021,Ryen2022}. 

The concept of photon emerges from the quantization of the classical electromagnetic field theory which can be carried through a systematic scheme consisting in the following steps: (a) Formulate Maxwell's equations in a Hamiltonian form, identifying a pair of canonical variables, defining a phase space such that the corresponding Hamilton equations are equivalent to Maxwell's equations. (b) Construct a complex Hilbert space $\mathcal{H}$ of classical configurations and a complex representation of the Hamiltonian structure. (c) Build a bosonic Fock space \cite{Berezin1966,Honegger2014} from the classical Hilbert space as 
\begin{eqnarray}
\mathbb{F}^{\mathfrak{B}}(\mathcal{H}):=\bigoplus_{n=0}^{\infty} \mathcal{H}^{\otimes_{S} n},
\end{eqnarray}
where $\mathcal{H}^{\otimes_S n}$ is $n$-times the symmetrized tensor product of $\mathcal{H}$ with itself. (d) Construct quantum observables from their classical equivalent using a correspondence principle. (e) Define the generator of the quantum dynamics in Fock space from the single-photon dynamics, which is determined by the classical Hamiltonian dynamics.
The choice of the classical Hamiltonian structure to build the quantized theory is thus central and allows to describe photons in different ways. The momentum description is often used since it diagonalizes the Hamiltonian of the system \cite{Mandel1995,CohenTannoudji1997,Loudon2000,Gerry2004,Garrison2008}; however, it does not provide a direct description of the photon states in position space, in particular for photon pulses. 

In this article, we will discuss two position space versions of the construction explained above and the relation between them. We will start with the standard quantization in the Coulomb gauge, using the vector potential and the electric field as canonical variables and the so called Landau-Peierls (LP) field as complex representation \cite{Landau1930,Cook1982}. Then we will introduce the quantization formulated by Bia{\l}ynicki-Birula (BB) \cite{BialynickiBirula1996,BialynickiBirula1998,BialynickiBirula2013} using the electric and magnetic fields as canonical variables and based on the Riemann-Silberstein complex vector \cite{Silberstein1907}. The main goal is to show that the two approaches are equivalent in the sense of Dirac's theory of transformations \cite{Dirac1989}, by showing that there is an isomorphism between their respective Hilbert spaces of states and between the observables. This implies that all the physical predictions are the same. We will construct the quantum theory in the Schr\"odinger picture, but once the model is set up, one can construct the corresponding Heisenberg and interaction pictures.
In the present formulation, we do not invoke the concept of photon wave function but only the concepts of states and observables, which are the two essential concepts needed in the quantum field theory of the electromagnetic field. Although the different possible representations are equivalent as we will show, they each have some advantages depending on the aspect of the theory one wants to look at. The BB quantization for instance does not require to choose a particular gauge which makes it a theoretically powerfull construction and it is also Lorentz invarient as it was shown in \cite{BialynickiBirula1996,Keller2005,Keller2014,Smith2007}. In particular, the BB representation is well adapted to study the electromagnetic energy in position space at the price of using some less usual tools in its contstruction such as a weighted scalar product. 
\vspace{.5cm}

All along the article, we will use the concept of helicity which is central in the construction of the BB classical Hamiltonian formalism and to make the link with the LP field. One can introduce this concept through the helicity operator $\Lambda$ defined by
\numparts
\begin{eqnarray}
\Lambda&:=(-\Delta)^{-1/2}\nabla\times\label{Lambda1}\\
&=(\nabla\times)(-\Delta)^{-1/2}\\
&=c\Omega^{-1}\nabla\times\label{helicity}\\
&={\rm sign}(\nabla\times)\label{Lambda4},
\end{eqnarray}
\endnumparts
where $\Delta=\nabla^2$ is the Laplace operator. We have used the notation
\begin{eqnarray}
\Omega:=c(-\Delta)^{1/2},
\end{eqnarray}
which denotes the frequency operator defined as the unique positive operator satisfying $\Omega^2=-c^2\Delta$ \cite{DeBievre2006,DeBievre2007,Federico2022}, we remark that $\left[\Lambda,\Omega\right]=0$. Since the curl operator can be written as
\begin{eqnarray}
\nabla\times = \left[\begin{array}{ccc}0 & -\partial_3 & \partial_2 \\ \partial_3 & 0 & -\partial_1 \\-\partial_2 & \partial_1 & 0\end{array}\right] = -i(\vec S \cdot \nabla),
\end{eqnarray}
with $\vec S  = (S^1,S^2,S^3)$, $S^j_{kl} = -i \epsilon_{jkl}$, and $\epsilon_{jkl}$ being the totally antisymmetric Levi-Civit\`a tensor, i.e.
\begin{eqnarray}
\fl 
S^1 = \left[\begin{array}{ccc}0 & 0 & 0 \\0 & 0 & -i\\0 & i & 0\end{array}\right], \qquad
S^2 = \left[\begin{array}{ccc}0 & 0 & i \\0 & 0 & 0 \\-i & 0 & 0\end{array}\right], \qquad
S^3 = \left[\begin{array}{ccc}0 & -i & 0 \\i & 0 & 0 \\0 & 0 & 0\end{array}\right].
\end{eqnarray}
We remark that the matrices $S^j$ are hermitian and they satisfy the spin-1 commutation relations $[ S^1,S^2] = i S^3$ (completed by cyclic permutations). Defining the operator $\vec{\hat p}:= -i\hbar \nabla$, in analogy with the momentum operator of quantum mechanics, one can write also $\nabla\times= (\vec S \cdot \vec{\hat p} )/\hbar$, $
 |\vec{\hat p}| =(\vec{\hat p}\cdot \vec{\hat p})^{1/2} = \hbar(-\Delta)^{1/2}$, and the helicity operator
\begin{eqnarray} \label{helicity-op-aa}
\Lambda = \vec S \cdot \frac{\vec{\hat p}}{|\vec{\hat p}|}.
\end{eqnarray}
This formula leads to the formal interpretation of the helicity operator as the projection of the spin $\vec S$ on the 
``direction of the motion'' defined by $\vec{\hat p}/|\vec{\hat p}|$. This interpretation is based on the analogy of these expressions with the ones of a quantum particle of spin 1. We emphasize however that at this stage we are dealing only with some operators, $\nabla\times$ and 
$\Omega$, that appear in complex representations of the classical Maxwell equations.

We will extensively use these operators to construct the LP field as well as to write the isomorphism making the link with the BB representation, which is the main result of this work. 
We remark that circular polarization plane waves 
\begin{equation}
\vec\phi_{\vec k,\pm}=\frac{1}{(2\pi)^{3/2}}\vec\epsilon_{\pm}(\vec k)e^{i\vec k\cdot\vec x},\label{plane waves}
\end{equation}
with
\begin{equation}
\vec\epsilon_+(\vec k)=\frac{1}{\sqrt{2}|\vec k|\sqrt{k_x^2+k_y^2}}\left(
\begin{array}{c}
-k_xk_z+i|\vec k|k_y\\
-k_yk_z-i|\vec k|k_x\\
k_x^2+k_y^2
\end{array}\right), \hspace{0.5cm}  \vec\epsilon_-(\vec k)=\vec\epsilon_+(\vec k)^\star ,
\end{equation}
are eigenfunctions of the three operators with eigenvalues $\nabla\times\vec\phi_{\vec k,\pm}=\pm|\vec k|\, \vec\phi_{\vec k,\pm}$, $\Lambda\vec\phi_{\vec k,\pm}=\pm\vec\phi_{\vec k,\pm}$ and $\Omega\vec\phi_{\vec k,\pm}=\omega_{\vec k}\vec\phi_{\vec k,\pm}$, where $\omega_{\vec k}=c|\vec k|>0$. Using the spectral theorem, one can properly define any power $\Omega^\alpha$ for a real number $\alpha\in[-2,2]$ by giving its action on the plane waves basis
\begin{equation}
\Omega^\alpha\vec\phi_{\vec k,\pm}=\omega^\alpha_{\vec k}\vec\phi_{\vec k,\pm}.
\end{equation}
The relations \eref{Lambda1}-\eref{Lambda4} can be shown by applying them to the basis functions \eref{plane waves}.
One can also show that any transverse field $\vec v$ can be decomposed into a sum of a positive and a negative helicity part $\vec v^{(h\pm)}$
\begin{eqnarray}
\vec v=\vec v^{(h+)}+\vec v^{(h-)},
\end{eqnarray}
where $\Lambda\vec v^{(h\pm)}=\pm\vec v^{(h\pm)}$. The positive and negative helicity parts can be constructed by applying the following projectors $\mathbb{P}^{(h\pm)}:=(1\pm\Lambda)/2$, i.e. $\vec v^{(h\pm)}=\mathbb{P}^{(h\pm)}\vec v$.

In the following, we will start by a summary of the canonical quantization of the free electromagnetic field in the Coulomb gauge using the LP field in position space and its equivalent in momentum space to make a link with the standard notations of the quantum optics literature \cite{Mandel1995,CohenTannoudji1997,Loudon2000,Gerry2004,Garrison2008}. Then we will introduce the BB quantization using a slightly different notation than the one used originaly by Bia{\l}ynicki-Birula, which is more convenient to write the isomorphism linking the LP and BB quantizations. After giving the explicit formulation of this isomorphism, we will show through a few examples that the two quantized theories are indeed equivalent and that the passage from one to the other is explicitely given by the isomorphism. Finally, we will discuss the time evolution within both theories, by explicitly linking their quantum Hamiltonian operators which represent the total energy of the system and the generator of the dynamics. This step ensures that the equivalence is preserved at any time.

\section{Canonical quantization in the Coulomb gauge}

Maxwell's equations in free space can be written in terms of the vector potential $\vec A$ in the Coulomb gauge as 
\numparts
\begin{eqnarray}
\frac{\partial^2\vec{A}}{\partial t^2}&=c^2\Delta\vec{A},\label{wave-equation}\label{Max eq}\\
\nabla\cdot\vec{A}&=0,\label{Max eq2}\\
\vec{B}&=\nabla\times \vec{A},\label{Max eq3}\\
\vec{E}&=-\frac{\partial\vec{A}}{\partial t},\label{Max eq4}
\end{eqnarray}
\endnumparts
with $c$ the speed of light in vaccum. In what follows, we will rewrite this set of equations to give it a Hamiltonian structure.

\subsection{Real canonical variables in position space}

By choosing as canonical variables $\vec{A}$ and its canonically conjugate variable $\vec{\Pi}=\varepsilon_0\frac{\partial\vec A}{\partial t}=-\varepsilon_0\vec{E}$, the wave equation (\ref{wave-equation}) can be written in Hamiltonian form
\numparts
\begin{eqnarray}
\frac{\partial\vec{A}}{\partial t}&=\frac{\delta H}{\delta \vec{\Pi}}=\frac{\vec{\Pi}}{\varepsilon_0},\label{h eq 1}\\
\frac{\partial\vec{\Pi}}{\partial t}&=-\frac{\delta H}{\delta \vec{A}}=\varepsilon_0c^2\Delta\vec{A}=-\varepsilon_0\Omega^2\vec{A},\label{h eq 2}
\end{eqnarray}
\endnumparts
where the Hamilton function $H$ is given by
\begin{eqnarray}
H&=\int_{\mathbb{R}^3} d^3x\left( \frac{1}{2\varepsilon_0} \vec{\Pi}\cdot \vec{\Pi}+\frac{\varepsilon_0}{2}\vec{A}\cdot \Omega^2\vec{A} \right).
\end{eqnarray}
The Hamilton equations (\ref{h eq 1}), (\ref{h eq 2}) and the constraints
\numparts
\begin{eqnarray}
\nabla\cdot\vec A&=0,\\
\nabla\cdot\vec\Pi&=0,
\end{eqnarray}
\endnumparts
are equivalent to Maxwell's equations \eref{Max eq}-\eref{Max eq4}.
\nosections

{\bf \noindent Remark:}
 The Hamilton function $H$ has a structure similar to a one-dimensional harmonic oscillator where $\varepsilon_0$ takes the place of the mass, and the operator $\Omega$ the place of the frequency. Because of this analogy, it has been stated that Maxwell's equations look like an infinite dimensional harmonic oscillator \cite{Mandel1995,CohenTannoudji1997,Loudon2000,Gerry2004,Garrison2008}. This image can be used as a guideline for the construction of the quantized theory in terms of a Fock space for the states.

\subsection{Complex representation of Maxwell's equations in position space --- The Landau-Peierls field}

We define the following complex field, called the Landau-Peierls (LP) field \cite{Landau1930,Cook1982,Cook1982a,Federico2022} as
\begin{eqnarray}
\vec{\psi}:=\frac{1}{\sqrt{2\hbar}} \left[ (\varepsilon_0 \Omega)^{1/2} \vec{A} +i (\varepsilon_0 \Omega)^{-1/2} \vec{\Pi}\right], \label{LPfunction}
\end{eqnarray}
in the Hilbert space 
\begin{eqnarray}
\mathcal{H}_{LP}:= \left\{\vec{\psi}(\vec{x})\Big| \nabla\cdot\vec{\psi}=0, \ \langle{\vec{\psi}}|\vec{\psi}\rangle_{LP}< \infty \right\},
\end{eqnarray}
with the scalar product
\begin{eqnarray}
\langle\vec{\psi}|\vec{\psi}'\rangle_{LP}:=\int_{\mathbb{R}^3} d^3 x\ \vec{\psi}^{\star}(\vec{x})\cdot\vec{\psi}'(\vec{x}).
\end{eqnarray}
The Hamilton function in terms of these complex variables takes the form
\begin{eqnarray}
H&=\hbar\int_{\mathbb{R}^3} d^3x\ \vec{\psi}^{\star}\cdot\Omega\vec{\psi},\label{HLP}
\end{eqnarray}
and the corresponding Hamilton equations are
\begin{equation}
i\frac{\partial\vec\psi}{\partial t}=\frac{1}{\hbar}\frac{\delta H}{\delta\vec\psi^\star}=\Omega\vec\psi.\label{max-LP}
\end{equation}
Together with the transversality constraint $\nabla\cdot\vec\psi=0$, they are equivalent to Maxwell's equations. We remark that equation \eref{max-LP} has the form of a Schr\"odinger equation where $\Omega$ is the generator of the dynamics.
Electromagnetic fields can be recovered from $\vec\psi$ by inverting (\ref{LPfunction}),
\numparts
\begin{eqnarray}
\vec A(\vec x)&=\sqrt{\frac{\hbar}{2\varepsilon_0}}\Omega^{-1/2}\left(\vec\psi(\vec x)+\vec\psi^\star(\vec x)\right),\label{A field}\\
\vec E(\vec x)&=i\sqrt{\frac{\hbar}{2\varepsilon_0}}\Omega^{1/2}\left(\vec\psi(\vec x)-\vec\psi^\star(\vec x)\right),\label{E field}\\
\vec B(\vec x)&=\sqrt{\frac{\hbar}{2\varepsilon_0}}\Omega^{-1/2}\nabla\times\left(\vec\psi(\vec x)+\vec\psi^\star(\vec x)\right).\label{B field}
\end{eqnarray}
\endnumparts
Using the definition of the helicity operator \eref{helicity}, one can write the inverse link between the Coulomb vector potential $\vec A$ and the magnetic field $\vec B=\nabla\times\vec A$ as
\begin{equation}
\vec A=c\Omega^{-1}\Lambda\vec B.
\end{equation}

\subsection{Momentum space formulation}

A similar procedure can be done in the reciprocal momentum (or Fourier) space. In this section we will only give the main results of that description and the link with the position description. For a more detailed overview of this formulation, we refer to the standard quantization of the free electromagnetic field of the quantum optics literature \cite{Mandel1995,CohenTannoudji1997,Loudon2000,Gerry2004,Garrison2008}.

We use the orthonormal basis of circular polarization plane waves \eref{plane waves} to decompose $\vec\psi$ as
\begin{eqnarray}
\vec\psi(\vec x)=\int_{\mathbb{R}^3}d^3k\sum_{\sigma=\pm}\vec\phi_{\vec k,\sigma}(\vec x)z(\vec k,\sigma),\label{inverse M}	
\end{eqnarray}
where $z(\vec k,\sigma):=\langle\vec\phi_{\vec k,\sigma}|\vec\psi\rangle_{LP}\in\mathbb{C}$. We define the momentum Hilbert space as
\begin{eqnarray}
\mathcal{H}_{m}:=\{z(\vec k,\sigma)\ |\ \langle z|z\rangle_{m}<\infty\},
\end{eqnarray}
endowed with the scalar product
\begin{eqnarray}
\langle z|z'\rangle_{m}:=\int_{\mathbb{R}^3}d^3k\sum_{\sigma=\pm}z^\star(\vec k,\sigma)z'(\vec k,\sigma).
\end{eqnarray}
The denomination of this space as momentum space is anticipated from the quantized theory since in that context, the variable $\vec k$ is proportionnal to the momentum of monochromatic photons \cite{Mandel1995,CohenTannoudji1997,Loudon2000,Gerry2004,Garrison2008}. It is also refered to as the Fourier space representation since it is related to the position space formulation by a decomposition into the circular plane waves basis which is very close to a Fourier transform. Indeed, one can pass from the position space description to the momentum description using the map
\begin{eqnarray}
\mathcal{M}:&\mathcal{H}_{LP}&\rightarrow\mathcal{H}_{m}\nonumber\\
&\vec\psi(\vec x)&\mapsto z(\vec k,\sigma)=\int_{\mathbb{R}^3}d^3x\ \vec\phi^\star_{\vec k,\sigma}(\vec x)\cdot\vec\psi(\vec x),\label{map M}
\end{eqnarray}
which is a bounded unitary symplectic transformation providing an isomorphism between the two Hilbert spaces. The inverse map $\mathcal{M}^{-1}$ is given by (\ref{inverse M}), and the equivalence of the scalar products reads
\begin{eqnarray}
\langle z|z'\rangle_{m}=\langle\mathcal{M}\psi|\mathcal{M}\vec\psi'\rangle_{m}=\langle \psi|\vec\psi'\rangle_{LP},\label{scalar prod eq}	
\end{eqnarray}
which is proven using the completeness of the basis $\{\vec\phi_{\vec k,\sigma}\}$.
The Hamilton function expressed in the new variables takes the form
\begin{eqnarray}
H=\int_{\mathbb{R}^3}d^3k\sum_{\sigma=\pm}\hbar\omega_{\vec k}\ z^\star(\vec k,\sigma)z(\vec k,\sigma),
\end{eqnarray}
and can be put in a harmonic oscillator form
\begin{eqnarray}\label{momentum-H}
H=\int_{\mathbb{R}^3} d^3k\sum_{\sigma=\pm}\left( \frac{1}{2\varepsilon_0} p^2_{\vec k,\sigma}+\frac{1}{2}\varepsilon_0\omega_{\vec k}^2 q^2_{\vec k,\sigma} \right),
\end{eqnarray}
by defining the real variables 
\numparts
\begin{eqnarray}
p_{\vec k,\sigma}&:=-i\sqrt{\frac{\hbar\varepsilon_0\omega_{\vec k}}{2}}(z-z^\star),\\
q_{\vec k,\sigma}&:=\sqrt{\frac{\hbar}{2\varepsilon_0\omega_{\vec k}}}(z+z^\star),\\
z(\vec k,\sigma)&=:\frac{1}{\sqrt{2\hbar}}\left((\varepsilon_0\omega_{\vec k})^{1/2}q_{\vec k,\sigma}+i(\varepsilon_0\omega_{\vec k})^{-1/2}p_{\vec k,\sigma}\right).
\end{eqnarray}
\endnumparts

\subsection{Quantization using a correspondence principle}

From the classical Hilbert spaces defined above, one can construct two bosonic Fock spaces defined by
\begin{eqnarray}
\mathbb{F}^{\mathfrak{B}}(\mathcal{H})=\bigoplus_{n=0}^{\infty}\mathcal{H}^{\otimes_S n}.
\end{eqnarray}
A strength of this construction is that $\mathcal{H}$ stands for either $\mathcal{H}_{LP}$ or $\mathcal{H}_{m}$. In the same spirit, the definition of creation-anihilation operators within the Fock space has the same form for both choices of the classical Hilbert space used:
\begin{eqnarray}
\hat B_{\eta} : \mathcal{H}^{\otimes_S l} \to \mathcal{H}^{\otimes_S (l-1)}, \qquad\qquad\qquad\hat B^\dag_{\eta} : \mathcal{H}^{\otimes_S l} \to \mathcal{H}^{\otimes_S (l+1)},\label{C-A operators}
\end{eqnarray}
with their action \cite{Berezin1966,Honegger2014} on the tensor product monomials defined by 
\numparts
\begin{eqnarray}
\hat B^\dag_{\eta} | \eta_1\otimes\ldots\otimes\eta_l \rangle & := &\sqrt{l+1}  ~\hat S_{l+1}| \eta\otimes \eta_1\otimes\ldots\otimes\eta_l \rangle, \label{Coperator}\\
\hat B_{\eta} | \eta_1\otimes\ldots\otimes\eta_l \rangle & := & ~\frac{1}{\sqrt{l} }\sum_{j=1}^l \langle\eta|\eta_j\rangle_{\mathcal{H}}~\hat S_{l-1}~ |  \eta_1\otimes\ldots\otimes\xcancel{\eta}_j\ldots\otimes\eta_l \rangle,\label{Aoperator}
\end{eqnarray}
\endnumparts
where $\eta_j$ are elements in $\mathcal{H}$, the notation $\xcancel{\eta}_j$ indicates that this term is missing and $\hat S_l$ are symmetrization operators. The bosonic creation-anihilation operators satisfy the following commutation relations \cite[equation (3.192)]{Garrison2008},\cite{Federico2022,Fabre2020}
\numparts
\begin{eqnarray} \label{commBBdag}
\left[  \hat B_{\eta_A}, \hat B^\dag_{\eta_B} \right]  & =  \langle {\eta_A} | {\eta_B} \rangle_{\mathcal{H}},  \\ 
\left[   \hat B_{\eta_A}, \hat B_{\eta_B} \right]  & =  0  =
\left[   \hat B^\dag_{\eta_A}, \hat B^\dag_{\eta_B} \right],\label{commutators}
\end{eqnarray}
\endnumparts
where $\langle \cdot | \cdot \rangle_{\mathcal{H}}$ denotes either $\langle \cdot | \cdot \rangle_{LP}$ or $\langle \cdot | \cdot \rangle_{m}$. The second ingredient to fully describe the quantum field theory is the representation of each observable by a selfadjoint operator acting on $\mathbb{F}^{\mathfrak{B}}(\mathcal{H})$. Which operators are to be associated to the physical observables have to be postulated. For a system like the Maxwell field that have a quadratic Hamiltonian, it can be done through a principle of correspondence, suggested by the fact that in the momentum space, the classical Hamilton function (\ref{momentum-H}) has the form of an infinite collection of independent harmonic oscillators.

\subsubsection{Quantization in the momentum space $\mathcal{H}_m$\\\\}

By this analogy, one postulates the following rule of correspondence, first defined in the momentum Fock space $\mathbb{F}^{\mathfrak{B}}(\mathcal{H}_{m})$:
\numparts
\begin{eqnarray}
{\rm Quantization~ map:}\hspace{1.5cm} &z(\vec k,\sigma)&\mapsto\hat B_{\varphi_{\vec k,\sigma}},\\
&z^\star(\vec k,\sigma)&\mapsto\hat B^\dag_{\varphi_{\vec k,\sigma}}.
\end{eqnarray}
\endnumparts
where $\varphi_{\vec k,\sigma}$ are the basis eigenfunctions in $\mathcal{H}_m$ given by
\numparts
\begin{eqnarray}
\vec\phi_{\vec k,\sigma}(\vec x)\mapsto\mathcal{M}\vec\phi_{\vec k,\sigma}&= \varphi_{\vec k,\sigma}(\vec k',\sigma')\\&=\delta_{\sigma,\sigma'}\delta(\vec k-\vec k').
\end{eqnarray}
\endnumparts

\subsubsection{Quantization in the Landau-Peierls space $\mathcal{H}_{LP}$\\\\}
By the isomorphism $\mathcal{M}$, the correspondence rule can be translated to the position LP Fock space $\mathbb{F}^{\mathfrak{B}}(\mathcal{H}_{LP})$ using (\ref{inverse M}):
\numparts
\begin{eqnarray}
{\rm Quantization~ map:}\hspace{1cm} &\vec\psi(\vec x)&\mapsto\int_{\mathbb{R}^3}d^3k\sum_{\sigma=\pm}\vec\phi_{\vec k,\sigma}(\vec x)\hat  B_{\vec\phi_{\vec k,\sigma}}=:\vec{\hat\Psi}(\vec x),\\
&\vec\psi^\star(\vec x)&\mapsto\int_{\mathbb{R}^3}d^3k\sum_{\sigma=\pm}\vec\phi^\star_{\vec k,\sigma}(\vec x)\hat  B^\dag_{\vec\phi_{\vec k,\sigma}}=:\vec{\hat\Psi}^\dag(\vec x),
\end{eqnarray}
\endnumparts
which defines the field operators $\vec{\hat\Psi}(\vec x)$ and $\vec{\hat\Psi}^\dag(\vec x)$. From these maps, one can write the electromagnetic field operators following \eref{A field}-\eref{B field}
\numparts
\begin{eqnarray}
\vec{\hat A}(\vec x)&=\sqrt{\frac{\hbar}{2\varepsilon_0}}\Omega^{-1/2}\left(\vec{\hat\Psi}(\vec x)+\vec{\hat\Psi}^\dag(\vec x)\right),\label{qA field}\\
\vec{\hat E}(\vec x)&=i\sqrt{\frac{\hbar}{2\varepsilon_0}}\Omega^{1/2}\left(\vec{\hat\Psi}(\vec x)-\vec{\hat\Psi}^\dag(\vec x)\right),\label{qE field}\\
\vec{\hat B}(\vec x)&=\sqrt{\frac{\hbar}{2\varepsilon_0}}\Omega^{-1/2}\nabla\times\left(\vec{\hat\Psi}(\vec x)+\vec{\hat\Psi}^\dag(\vec x)\right).\label{qB field}
\end{eqnarray}
\endnumparts
\nosections

{\bf \noindent Remark:} We point out that the interpretation of the field operators $\vec{\hat\Psi}^\dag$ and $\vec{\hat\Psi}$ is not to be confused with that of the creation-anihilation operators $\hat  B^\dag_{\vec\psi}$ and $\hat  B_{\vec\psi}$ for $\vec\psi\in\mathcal{H}_{LP}$. Indeed, when applied to the vacuum state, $\hat  B^\dag_{\vec\psi}$ creates a photon carried by the classical solution $\vec\psi$ of Maxwell's equations \cite{Federico2022}, while each component $\hat\Psi_j^\dag$ is an operator-valued distribution which has to be integrated over the whole space acting on a test function to properly create a photon.

\subsection{Equivalence between the LP and momentum quantizations}
\label{eq LP-m}

In this section, we show that the isomorphism $\mathcal{M}$ given by equations \eref{inverse M}, \eref{map M} and \eref{scalar prod eq}, provides a direct relation between the quantized theories we have constructed above. Indeed, since the central concept of bosonic creation-anihilation operators, which allows one to write any state and observable in the theory, are explicitely linked by the classical isomorphism $\mathcal{M}$, all physical predictions made with the two representations are the same. 

If one considers e.g. a single-photon state in the position space LP representation which reads as
\begin{eqnarray}
\hat B_{\vec\psi}^\dag|\varnothing\rangle_{LP}=|\vec\psi\rangle_{LP},
\end{eqnarray}
with $\vec\psi\in\mathcal{H}_{LP}$. One can use the isomorphism $\mathcal{M}$ to write it in terms of a creation operator in the momentum space as follows 
\numparts
\begin{eqnarray}
|\vec \psi\rangle_{LP}&=|\mathcal{M}^{-1}z\rangle_{LP}\\
&=\mathcal{M}^{-1}|z\rangle_{m}\\
&=\mathcal{M}^{-1}\hat B_z^\dag|\varnothing\rangle_{m}\\
&=\mathcal{M}^{-1}\hat B_z^\dag \mathcal{M}|\varnothing\rangle_{LP},
\end{eqnarray}
\endnumparts
from which we deduce that $\hat B_{\vec\psi}^\dag=\mathcal{M}^{-1}\hat B_z^\dag \mathcal{M}=\mathcal{M}^{-1}\hat B_{\mathcal{M}\vec\psi}^\dag \mathcal{M}$, where $\hat B_z^\dag: \mathcal{H}_{m}^{\otimes_S l} \to \mathcal{H}_{m}^{\otimes_S (l+1)}$ is the creation operator acting on the momentum Fock space $\mathbb{F}^{\mathfrak{B}}(\mathcal{H}_{m})$. We have extended here the isomorphism $\mathcal{M}$ to the whole Fock space by defining its action on the vacuum state: $\mathcal{M}|\varnothing\rangle_{LP}=|\varnothing\rangle_m$. The same state can then be described in one or the other representation without ambiguity. 
Regarding the observables, one can for instance express the electric field in terms of creation-anihilation operators in the LP position representation as
\numparts
\begin{eqnarray}
\fl
\vec{\hat E}_{LP}(\vec x)&=i\sqrt{\frac{\hbar}{2\varepsilon_0}}\Omega^{1/2}\left(\vec{\hat\Psi}(\vec x)-\vec{\hat\Psi}^\dag(\vec x)\right)\\
\fl
&=i\sqrt{\frac{\hbar}{2\varepsilon_0}}\int_{\mathbb{R}^3}d^3k\sum_{\sigma=\pm}\omega_{\vec k}^{1/2}\left(\vec\phi_{\vec k,\sigma}(\vec x)\hat  B_{\vec\phi_{\vec k,\sigma}}-\vec\phi^\star_{\vec k,\sigma}(\vec x)\hat  B^\dag_{\vec\phi_{\vec k,\sigma}}\right),
\end{eqnarray}
\endnumparts
which is expressed in the momentum representation by
\numparts
\begin{eqnarray}
\fl
\vec{\hat E}_{m}(\vec x)&=\mathcal{M}\vec{\hat E}_{LP}\mathcal{M}^{-1}\\
\fl
&=i\sqrt{\frac{\hbar}{2\varepsilon_0}}\int_{\mathbb{R}^3}d^3k\sum_{\sigma=\pm}\omega_{\vec k}^{1/2}\left(\vec\phi_{\vec k,\sigma}(\vec x)\mathcal{M}\hat  B_{\vec\phi_{\vec k,\sigma}}\mathcal{M}^{-1}-\vec\phi^\star_{\vec k,\sigma}(\vec x)\mathcal{M}\hat  B^\dag_{\vec\phi_{\vec k,\sigma}}\mathcal{M}^{-1}\right) \\
\fl
&=i\sqrt{\frac{\hbar}{2\varepsilon_0}}\int_{\mathbb{R}^3}d^3k\sum_{\sigma=\pm}\omega_{\vec k}^{1/2}\left(\vec\phi_{\vec k,\sigma}(\vec x)\hat  B_{\varphi_{\vec k,\sigma}}-\vec\phi^\star_{\vec k,\sigma}(\vec x)\hat  B^\dag_{\varphi_{\vec k,\sigma}}\right)\\
\fl
&=i\sqrt{\frac{\hbar}{2\varepsilon_0}}\int_{\mathbb{R}^3}d^3k\sum_{\sigma=\pm}\omega_{\vec k}^{1/2}\left(\vec\phi_{\vec k,\sigma}(\vec x)\hat  a_{\vec k,\sigma}-\vec\phi^\star_{\vec k,\sigma}(\vec x)\hat  a^\dag_{\vec k,\sigma}\right),
\end{eqnarray}
\endnumparts 
where we have used in the last equality the usual notation $\hat a_{\vec k,\sigma}=\hat B_{\delta_{\sigma,\sigma'}\delta(\vec k-\vec k')}$ and $\hat a^\dag_{\vec k,\sigma}=\hat B^\dag_{\delta_{\sigma,\sigma'}\delta(\vec k-\vec k')}$ \cite{Mandel1995,CohenTannoudji1997,Loudon2000,Gerry2004,Garrison2008}. Furthemore, if one plugs in the expression of the circular plane wave eigenfunctions \eref{plane waves}, we obtain the usual expression of the quantum optics literature \cite{Mandel1995,CohenTannoudji1997,Loudon2000,Gerry2004,Garrison2008}
\begin{eqnarray}
\fl
\vec{\hat E}_{m}(\vec x)&=i\sqrt{\frac{\hbar}{2\varepsilon_0(2\pi)^3}}\int_{\mathbb{R}^3}d^3k\sum_{\sigma=\pm}\omega_{\vec k}^{1/2}\left(\vec\epsilon_	\sigma(\vec k) e^{i\vec k\cdot\vec x}\hat  a_{\vec k,\sigma}-\vec\epsilon_\sigma(\vec k)^\star e^{-i\vec k\cdot\vec x}\hat  a^\dag_{\vec k,\sigma}\right).
\end{eqnarray}
With this example, we have shown how one can pass from one representation to the other using directly the classical isomorphism $\mathcal{M}$ in a standard way, i.e. directly applied to the states and through a similarity relation for creation-anihilation operators or observables. This simple relation guarantees that the predictions one can make with one theory is consistent with what is obtained with the other.

\section{Bia{\l}ynicki-Birula's complex formulation of Maxwell's equations --- Choice of the canonical variables}

We consider now another approach for the quantization which has been developped mainly by Bia{\l}ynicki-Birula \cite{BialynickiBirula1996,BialynickiBirula1998,BialynickiBirula2013}. One of the advantages of this construction is that it does not require to choose any particular gauge since it starts by defining the  canonical variables to be directly proportional to the electric and magnetic fields as \cite[p.25, Section 10]{BialynickiBirula2013}
\numparts
\begin{eqnarray}
\vec\mathcal{P}_{RS}&:=\frac{\vec B}{\sqrt{\mu_0}},\\
\vec\mathcal{Q}_{RS}&:=\sqrt{\varepsilon_0}\vec E.
\end{eqnarray}
\endnumparts
The Hamilton function is defined by
\numparts
\begin{eqnarray}
K_{RS}&:=\frac{c}{2}\int_{\mathbb{R}^3}d^3x\left( \vec\mathcal{P}_{RS}\cdot\nabla\times\vec\mathcal{P}_{RS}  +  \vec\mathcal{Q}_{RS}\cdot\nabla\times\vec\mathcal{Q}_{RS} \right)\\
&=\frac{c}{2}\int_{\mathbb{R}^3}d^3x\left( \varepsilon_0\vec E\cdot\nabla\times\vec E  +  \frac{1}{\mu_0}\vec B\cdot\nabla\times\vec B \right),
\end{eqnarray}
\endnumparts
yielding the following Hamilton equations
\numparts
\begin{eqnarray}
\frac{\partial\vec\mathcal{P}_{RS}}{\partial t}&=-\frac{\delta K_{RS}}{\delta\vec\mathcal{Q}_{RS}}=-c\nabla\times\vec\mathcal{Q}_{RS},\\
\frac{\partial\vec\mathcal{Q}_{RS}}{\partial t}&=\frac{\delta K_{RS}}{\delta\vec\mathcal{P}_{RS}}=c\nabla\times\vec\mathcal{P}_{RS},
\end{eqnarray}
\endnumparts
that are equivalent to Maxwell's equations. We denote the Hamilton function by $K_{RS}$ and not by $H$, to emphasize that it is not the total electromagnetic energy \cite{CohenTannoudji1997,Jackson1998}, which reads
\numparts
\begin{eqnarray}
\mathcal{E}_{tot}&=\frac{1}{2}\int_{\mathbb{R}^3}d^3x\left( \varepsilon_0\vec E\cdot\vec E +\frac{1}{\mu_0}\vec B\cdot\vec B \right)\\
&=\frac{1}{2}\int_{\mathbb{R}^3}d^3x\left( \vec\mathcal{P}_{RS}\cdot\vec\mathcal{P}_{RS} +  \vec\mathcal{Q}_{RS}\cdot \vec\mathcal{Q}_{RS} \right).
\end{eqnarray}
\endnumparts

\subsection{Complex representation of Maxwell's equations in position space --- the Riemann-Silberstein vector}

From the canonical variables defined above one can construct a complex representation of the field called the Riemann-Silberstein (RS) vector \cite{BialynickiBirula2013,Silberstein1907}
\begin{eqnarray}
\vec F_{RS}:=\frac{1}{\sqrt{2}}\left( \vec\mathcal{Q}_{RS}+i\vec\mathcal{P}_{RS} \right)=\sqrt{\frac{\varepsilon_0}{2}}\left(\vec E+ic\vec B\right).
\end{eqnarray}
Maxwell's equations can be written in the RS complex representation as 
\begin{eqnarray}
i\frac{\partial\vec F_{RS}}{\partial t}=-\frac{\delta K_{RS}}{\delta\vec F_{RS}^\star}=c\nabla\times\vec F_{RS},
\end{eqnarray}
with
\begin{eqnarray}
K_{RS}=c\int_{\mathbb{R}^3}d^3x\ \vec F_{RS}^\star\cdot\nabla\times\vec F_{RS}.
\end{eqnarray}
In analogy with what we did for the Coulomb gauge, we define the symplectic unitary map
\begin{eqnarray}
\vec F_{RS}\mapsto z_{RS}(\vec k, \sigma)=\int_{\mathbb{R}^3}d^3x\ \vec\phi_{\vec k,\sigma}^\star(\vec x)\cdot\vec F_{RS}(\vec x),
\end{eqnarray}
from which we deduce the Hamilton function in terms of the RS momentum variable $z_{RS}$
\begin{eqnarray}
K_{RS}=\int_{\mathbb{R}^3}d^3k\left[ \omega_{\vec k}\ z_{RS}^\star(\vec k,+)z_{RS}(\vec k,+) - \omega_{\vec k}\ z_{RS}^\star(\vec k,-)z_{RS}(\vec k,-) \right].
\end{eqnarray}
Written in terms of the real variables defined as
\numparts
\begin{eqnarray}
p^{RS}_{\vec k,\sigma}&:=-i\sqrt{\frac{\varepsilon_0\omega_{\vec k}}{2}}(z_{RS}-z_{RS}^\star),\\
q^{RS}_{\vec k,\sigma}&:=\sqrt{\frac{1}{2\varepsilon_0\omega_{\vec k}}}(z_{RS}+z_{RS}^\star),\\
z_{RS}(\vec k,\sigma)&=:\frac{1}{\sqrt{2}}\left((\varepsilon_0\omega_{\vec k})^{1/2}q^{RS}_{\vec k,\sigma}+i(\varepsilon_0\omega_{\vec k})^{-1/2}p^{RS}_{\vec k,\sigma}\right),
\end{eqnarray}
\endnumparts
the Hamilton function becomes
\begin{eqnarray}
\fl
K_{RS}=\int_{\mathbb{R}^3}d^3k\left[ \left( \frac{(p_{\vec k,+}^{RS})^2}{2\varepsilon_0} +\frac{\varepsilon_0}{2}\omega^2_{\vec k} (q^{RS}_{\vec k, +})^2\right) - \left( \frac{(p^{RS}_{\vec k,-})^2}{2\varepsilon_0} +\frac{\varepsilon_0}{2}\omega^2_{\vec k} (q^{RS}_{\vec k, -})^2\right) \right].
\end{eqnarray}
We see here that it looks like two infinite collections of independent harmonic oscillators but the second one appears with a minus sign. This situation is similar to the one encountered with the Dirac equation, in which the electrons could have positive and negative energies. It poses a difficulty for the construction of the quantized model. However, this difficulty can be avoided by making a different choice for the canonical variables, as we describe in the next section. The construction we are going to introduce is strongly inspired by the work of Bia{\l}ynicki-Birula (BB) \cite{BialynickiBirula1996} which is the reason why we will call it the BB complex representation, although we formulate it in a slightly different form.

\subsection{Alternative choice of canonical variables that avoid the negative eigenvalues in the Hamiltonian}

To construct canonical variables that avoid the difficulty of the negative eigenvalues describe above, we have to locate the variables that lead to negative contributions. This can be done using the notion of helicity which allows us to decompose the RS canonical variables into their positive and negative helicity parts $\vec F_{RS}=\vec F_{RS}^{(h+)}+\vec F_{RS}^{(h-)}$ or equivalently in terms of the real variables $\vec\mathcal{P}_{RS}=\vec\mathcal{P}_{RS}^{(h+)}+\vec\mathcal{P}_{RS}^{(h-)}$ and $\vec\mathcal{Q}_{RS}=\vec\mathcal{Q}_{RS}^{(h+)}+\vec\mathcal{Q}_{RS}^{(h-)}$, in terms of which the Hamilton function takes the form
\numparts
\begin{eqnarray}
K_{RS}&=\frac{c}{2}\int_{\mathbb{R}^3}d^3x\left[ \left(\vec\mathcal{P}_{RS}^{(h+)}\cdot\nabla\times\vec\mathcal{P}_{RS}^{(h+)}  +  \vec\mathcal{Q}_{RS}^{(h+)}\cdot\nabla\times\vec\mathcal{Q}_{RS}^{(h+)} \right) \right.\nonumber\\ &\:\:\:\:\:\:\:\:\:\:\:\:\:\:\:\:\:\:\:\:\: \left.+ \left(\vec\mathcal{P}_{RS}^{(h-)}\cdot\nabla\times\vec\mathcal{P}^{(h-)}  +  \vec\mathcal{Q}_{RS}^{(h-)}\cdot\nabla\times\vec\mathcal{Q}_{RS}^{(h-)} \right) \right] \\
&=\frac{1}{2}\int_{\mathbb{R}^3}d^3x\left[ \left(\vec\mathcal{P}_{RS}^{(h+)}\cdot\Omega\vec\mathcal{P}_{RS}^{(h+)}  +  \vec\mathcal{Q}_{RS}^{(h+)}\cdot\Omega\vec\mathcal{Q}_{RS}^{(h+)} \right) \right.\nonumber\\ &\:\:\:\:\:\:\:\:\:\:\:\:\:\:\:\:\:\:\:\:\: \left. - \left(\vec\mathcal{P}_{RS}^{(h-)}\cdot\Omega\vec\mathcal{P}_{RS}^{(h-)}  +  \vec\mathcal{Q}_{RS}^{(h-)}\cdot\Omega\vec\mathcal{Q}_{RS}^{(h-)} \right) \right],
\end{eqnarray}
\endnumparts
where we have used $c\nabla\times=\Omega\Lambda$ to obtain the second expression and the fact that $\Lambda\vec v^{(h\pm)}=\pm\vec v^{(h\pm)}$. It shows that the negative contributions identified in the momentum representation come from the negative helicity part of the field. To circumvent this problem, one can define new canonical variables that we call the BB variables
\begin{eqnarray}
\vec \mathcal{Q}_{BB}&:=\vec \mathcal{Q}_{RS},\hspace{2cm} \vec \mathcal{P}_{BB}&:=\Lambda\vec \mathcal{P}_{RS},\label{BB canonical var}
\end{eqnarray}
and a new classical Hamilton function
\begin{eqnarray}
K_{BB}&:=\frac{1}{2}\int_{\mathbb{R}^3}d^3x\left[ \left(\vec\mathcal{P}_{BB}^{(h+)}\cdot\Omega\vec\mathcal{P}_{BB}^{(h+)}  +  \vec\mathcal{Q}_{BB}^{(h+)}\cdot\Omega\vec\mathcal{Q}_{BB}^{(h+)} \right) \right.\nonumber\\ &\:\:\:\:\:\:\:\:\:\:\:\:\:\:\:\:\:\:\:\:\: \left. + \left(\vec\mathcal{P}_{BB}^{(h-)}\cdot\Omega\vec\mathcal{P}_{BB}^{(h-)}  +  \vec\mathcal{Q}_{BB}^{(h-)}\cdot\Omega\vec\mathcal{Q}_{BB}^{(h-)} \right) \right],\label{KBB}
\end{eqnarray}
which is positive since $\Omega$ is a positive operator, as opposed to $\nabla\times$. This alternative choice of the canonical variables thus avoids the problem of negative eigenvalues.
We emphasize that $K_{BB}$ is not the Hamilton function $K_{RS}$ expressed in the new variables since the relation \eref{BB canonical var} is not a canonical transformation. 

\subsection{The Bia{\l}ynicki-Birula complex representation}

We choose a different Hamiltonian structure that gives the same classical Maxwell equations.
For the complex representation, we define a modification of the RS vector adapted to the new canonical variables which we call the Bia{\l}ynicki-Birula vector
\begin{eqnarray}
\vec F_{BB}:=\frac{1}{\sqrt{2}}\left(\vec \mathcal{Q}_{BB}+i\vec \mathcal{P}_{BB}\right)=\sqrt{\frac{\varepsilon_0}{2}}\left(\vec E+ic\Lambda\vec B\right),\label{FBB}
\end{eqnarray}
which can be decomposed in the six-component field (bispinor) notation used originally in \cite{BialynickiBirula1996}
\begin{equation}
\Psi_{BB}:=\left(
\begin{array}{c}
\vec F_{BB}^{(h+)} \\
\vec F_{BB}^{(h-)} \\
\end{array}\right)=\left(
\begin{array}{c}
\vec F_{RS}^{(h+)} \\
\vec F_{RS}^{(h-)\star} \\
\end{array}\right).
\end{equation}
\nosections

{\bf \noindent Remarks:} \begin{itemize}

\item The BB vector $\vec F_{BB}$ that we introduced here was not mentionned explicitely in \cite{BialynickiBirula1996,BialynickiBirula2013} but is a completely equivalent formulation of the original formulation using the six-component field $\Psi_{BB}$. We prefer, however, to use $\vec F_{BB}$ since it is a more compact notation which allows us to write simpler formulas especially when we will introduce the isomorphism with the LP field.

\item 
In the original works of BB \cite{BialynickiBirula1996}, the six-component field $\Psi_{BB}$ was defined using the positive and negative frequency parts of the RS field which we denote by $\vec F_{RS}^{(f\pm)}$, instead of the positive and negative helicity parts as we have done here. This difference does not change anything for the definition of $\Psi_{BB}$ since $\vec F_{RS}^{(h\pm)}=\vec F_{RS}^{(f\pm)}$. However, we stress out that this property is in general not true, e.g. $\vec \psi^{(h\pm)}\neq\vec\psi^{(f\pm)}$ since $\vec\psi^{(f-)}=0$ for any $\vec\psi\in\mathcal{H}_{LP}$ while $\vec\psi^{(h-)}\neq0$ in general. Other examples are the electric and the magnetic fields for which $\vec E^{(h\pm)}\neq\vec E^{(f\pm)}$ and $\vec B^{(h\pm)}\neq\vec B^{(f\pm)}$.

\item 
The structure of the Bia{\l}ynicki-Birula six-component spinor can be illustrated by the description of the general elliptically polarized plane waves propagating in the $z$-direction, i.e. $\vec k=(0\ 0\ k)^T$ with $k>0$. The electric $\vec E^\pm$ and the magnetic $\vec B^\pm$ fields of a left/right cirlcularly polarized plane wave propagating in direction $z$ can be written as \cite[p.299, eq. (7.21)]{Jackson1998}
\numparts
\begin{eqnarray}
\vec E^\pm&=&\frac{1}{\sqrt{\varepsilon_0(2\pi)^3}}\left(
\begin{array}{c}
\cos(kz-\omega t)\\\mp\sin(kz-\omega t)\\0
\end{array}\right),\\
\vec B^\pm&=&\frac{1}{c\sqrt{\varepsilon_0(2\pi)^3}}\left(
\begin{array}{c}
\pm\sin(kz-\omega t)\\\cos(kz-\omega t)\\0
\end{array}\right).
\end{eqnarray}
\endnumparts
We remark that the left/right circularly polarized plane waves have respectively $+/-$ helicity, since they are eigenfunctions of the curl operator $\nabla\times$ with positive and negative eigenvalues.

\noindent
The plane wave with left circular polarization is represented by the Bia{\l}ynicki-Birula six-component spinor
\begin{eqnarray}
\fl
\Psi_{BB}^{\rm (left~circular)}=\sqrt{\frac{\varepsilon_0}{2}}\left(
\begin{array}{c}
\vec E^++ic\vec B^+\\
\vec 0
\end{array}\right)=(2\pi)^{-3/2}e^{i(kz-\omega t)}\left(
\begin{array}{c}
\vec \epsilon_+\\
\vec 0
\end{array}\right),
\end{eqnarray}
with $\vec\epsilon_+=\frac{1}{\sqrt{2}}\left(1\ i \ 0\right)^T$. The plane wave with right circular polarization is represented by
\begin{eqnarray}
\fl
\Psi_{BB}^{\rm (right~circular)}=\sqrt{\frac{\varepsilon_0}{2}}\left(
\begin{array}{c}
\vec 0\\
\vec E^--ic\vec B^-
\end{array}\right)=(2\pi)^{-3/2}e^{i(kz-\omega t)}\left(
\begin{array}{c}
\vec 0\\
\vec \epsilon_-
\end{array}\right),
\end{eqnarray}
with $\vec\epsilon_-=\frac{1}{\sqrt{2}}\left(1\ -i\ 0\right)^T$.

\noindent
The single-photon plane wave with general elliptic polarization is represented by
the linear combinations
\begin{eqnarray}
\Psi_{BB}^{\rm (elliptic)}&=&\alpha_+\Psi_{BB}^{\rm (left~circular)}+\alpha_-\Psi_{BB}^{\rm (right~circular)}\\
&=&(2\pi)^{-3/2}e^{i(kz-\omega t)}\left(\begin{array}{c}\alpha_+\vec\epsilon_+\\\alpha_-\vec\epsilon_-\end{array}\right),
\end{eqnarray}
where $\alpha_\pm\in\mathbb{C}$, $|\alpha_+|^2+|\alpha_-|^2=1$.

\end{itemize}
\nosections

\noindent The Hamilton function $K_{BB}$ in the BB complex representation reads
\begin{eqnarray}
K_{BB}&=\int_{\mathbb{R}^3}d^3x\ \vec F_{BB}^\star\cdot\Omega\vec{F}_{BB},
\end{eqnarray}
and the corresponding Hamilton equations can be written as 
\begin{equation}
i\frac{\partial \vec F_{BB}}{\partial t}=\frac{\delta K_{BB}}{\delta\vec F_{BB}^\star}=\Omega\vec{F}_{BB}.\label{time ev FBB}
\end{equation}
Using the six-component notation $\Psi_{BB}$ the last two equations are equivalent to 
\begin{eqnarray}
K_{BB}=\int_{\mathbb{R}^3}d^3x\ \Psi_{BB}^\star\cdot\Omega\Psi_{BB},
\end{eqnarray}
and
\begin{equation}
i\frac{\partial \Psi_{BB}}{\partial t}=\frac{\delta K_{BB}}{\delta\Psi_{BB}^\star}=\Omega\Psi_{BB}.
\label{time ev calFBB}
\end{equation}
\nosections

{\bf \noindent Remark:} The construction introduced here is similar to what is done for the Dirac equation to remove the negative energy solutions and which led to the interpretation of anti-matter. However, in our case the negative contributions are not negative energies since $K_{RS}$ is not the energy of the field. We remove them because it is more convenient for the quantization as we will see later, and it allows to make the link with the quantization in the Coulomb gauge explicit.

\subsection{Quantization in the Bia{\l}ynicki-Birula representation}

We associate to the BB vector, a classical Hilbert space $\mathcal{H}_{BB}$ defined as
\begin{eqnarray}
\mathcal{H}_{BB}&:= \left\{\vec F_{BB}(\vec{x})\Big| \langle\vec F_{BB}|\vec F_{BB}\rangle_{BB}< \infty \right\}\label{H BB}
\end{eqnarray}
with the following weighted scalar product \cite{BialynickiBirula1996,BialynickiBirula1998,BialynickiBirula2013}
\begin{eqnarray}
\langle\vec F_{BB}|\vec F'_{BB}\rangle_{BB}=\frac{1}{\hbar}\int_{\mathbb{R}^3}d^3x\ \vec F_{BB}(\vec x)\cdot\Omega^{-1}\vec F'_{BB}(\vec x).
\end{eqnarray}
The main motivation for the choice of this weighted scalar product is that it is Lorentz invariant \cite{BialynickiBirula1996,Keller2005,Smith2007,Keller2014}. We will see that it is also an essential ingredient for the isomorphism with the standard quantization in the Coulomb gauge.

\nosections
{\bf \noindent Remark:} Fields with different helicities are orthogonal to each other with respect to both scalar products $\langle\cdot|\cdot\rangle_{LP}$ and $\langle\cdot|\cdot\rangle_{BB}$.
\nosections

\noindent
Classical electromagnetic fields can be expressed in terms of the BB vector by inverting equation \eref{FBB} which gives
\numparts
\begin{eqnarray}
\vec E(\vec x)&=\frac{1}{\sqrt{2\varepsilon_0}}\left(\vec F_{BB}(\vec x)+\vec F_{BB}^\star(\vec x)\right),\\
\vec B(\vec x)&=\frac{-i}{\sqrt{2\varepsilon_0c^2}}\Lambda\left(\vec F_{BB}(\vec x)-\vec F_{BB}^\star(\vec x)\right).
\end{eqnarray}
\endnumparts
From this classical Hilbert space $\mathcal{H}_{BB}$, one can construct the bosonic Fock space $\mathbb{F}^{\mathfrak{B}}(\mathcal{H}_{BB})$ by the general procedure \eref{C-A operators}-\eref{commutators}. The quantum observables associated to the classical physical quantities are then obtained by the following correspondence principle in analogy with what was done for the Coulomb gauge and with the harmonic oscillator structure of the Hamiton function. Indeed, one can develop the BB vector into
\numparts
\begin{eqnarray}
\vec F_{BB}(\vec x)&=\int_{\mathbb{R}^3}d^3k\sum_{\sigma=\pm}\vec g_{\vec k,\sigma}(\vec x)z_{BB}(\vec k,\sigma),\;\;\;\; z_{BB}(\vec k,\sigma)=\langle\vec g_{\vec k,\sigma}|\vec F_{BB}\rangle_{BB},\label{decompostition FBB}\\
\vec F^\star_{BB}(\vec x)&=\int_{\mathbb{R}^3}d^3k\sum_{\sigma=\pm}\vec g^\star_{\vec k,\sigma}(\vec x)z^\star_{BB}(\vec k,\sigma),\label{decompostition FBB2}
\end{eqnarray}
\endnumparts
where the functions $\{\vec g_{\vec k,\sigma}\}$ are an orthonormal basis of $\mathcal{H}_{BB}$, which can be taken as $\vec g_{\vec k,\sigma}=i\sqrt{\hbar\omega_{\vec k}}\vec\phi_{\vec k,\sigma}$. The quantization map of the BB representation is thus defined by 
\numparts
\label{BBquantization}
\begin{eqnarray}
z_{BB}&\mapsto\hat  C_{\vec g_{\vec k,\sigma}},\label{BBquantization1}\\
z^\star_{BB}&\mapsto\hat  C^\dag_{\vec g_{\vec k,\sigma}}, \label{BBquantization2}
\end{eqnarray}
\endnumparts
where $\hat  C_{\vec g_{\vec k,\sigma}}$ and $\hat  C^\dag_{\vec g_{\vec k,\sigma}}$ are creation-anihilation operators in the BB Fock space $\mathbb{F}^{\mathfrak{B}}(\mathcal{H}_{BB})$, defined by the general construction \eref{Coperator} and \eref{Aoperator}. We define also BB field operators as
\numparts
\begin{eqnarray}
\vec{\hat F}_{BB}(\vec x)&:=\int_{\mathbb{R}^3}d^3k\sum_{\sigma=\pm}\vec g_{\vec k,\sigma}(\vec x)\hat  C_{\vec g_{\vec k,\sigma}},\label{BB field op1}\\
\vec{\hat F}^\dag_{BB}(\vec x)&:=\int_{\mathbb{R}^3}d^3k\sum_{\sigma=\pm}\vec g^\star_{\vec k,\sigma}(\vec x)\hat  C^\dag_{\vec g_{\vec k,\sigma}}.\label{BB field op2}
\end{eqnarray}
\endnumparts
Quantized electromagnetic observables are given directly by the operators
\numparts
\begin{eqnarray}
\label{E field BB}
\vec{\hat E}(\vec x)&=\frac{1}{\sqrt{2\varepsilon_0}}\left(\vec{\hat F}_{BB}(\vec x)+\vec{\hat F}^\dag_{BB}(\vec x)\right),\\
\vec{\hat B}(\vec x)&=\frac{-i}{\sqrt{2\varepsilon_0c^2}}\Lambda\left(\vec{\hat F}_{BB}(\vec x)-\vec{\hat F}^\dag_{BB}(\vec x)\right).
\end{eqnarray}
\endnumparts
One can check that it corresponds to the quantization proposed by BB, as written e.g. in \cite{BialynickiBirula1998} by computing the form of the RS field operator: 
\numparts
\begin{eqnarray}
\vec{\hat F}_{RS}&=\sqrt{\frac{\varepsilon_0}{2}}\left( \vec{\hat E} +ic\vec{\hat B} \right) =\vec{\hat F}_{BB}^{(h+)}+\vec{\hat F}_{BB}^{(h-)\dag}\\
&=\int d^3k\sum_{\sigma=\pm}\left[\left( \frac{1+\Lambda}{2}\right)\vec g_{\vec k,\sigma}(\vec x)\hat  C_{\vec g_{\vec k,\sigma}}+\left( \frac{1-\Lambda}{2}\right)\vec g_{\vec k,\sigma}^\star(\vec x)\hat  C^\dag_{\vec g_{\vec k,\sigma}}\right]\\
&=\int d^3k\sum_{\sigma=\pm}\left[\mathbb{P}^{(h+)}\vec g_{\vec k,\sigma}(\vec x)\hat  C_{\vec g_{\vec k,\sigma}}+\mathbb{P}^{(h-)}\vec g_{\vec k,\sigma}^\star(\vec x)\hat  C^\dag_{\vec g_{\vec k,\sigma}}\right]\\
&=\int d^3k\left[\vec g_{\vec k,+}(\vec x)\hat  C_{\vec g_{\vec k,+}}+\vec g_{\vec k,-}^\star(\vec x)\hat  C^\dag_{\vec g_{\vec k,-}}\right]\\
&=\int d^3k\left[i\sqrt{\frac{\hbar\omega_{\vec k}}{(2\pi)^3}}\vec\epsilon_+(\vec k)e^{i\vec k\cdot\vec x}\hat  C_{\vec g_{\vec k,+}}+i\sqrt{\frac{\hbar\omega_{\vec k}}{(2\pi)^3}}\vec\epsilon_-(\vec k)^\star e^{-i\vec k\cdot\vec x}\hat  C^\dag_{\vec g_{\vec k,-}}\right]\\
&=\int d^3k\sqrt{\frac{\hbar\omega_{\vec k}}{(2\pi)^3}}\vec e(\vec k)\left[e^{i\vec k\cdot\vec x}\hat  a(\vec k)+e^{-i\vec k\cdot\vec x}\hat  b^\dag(\vec k)\right],\label{F_RS BB}
\end{eqnarray}
\endnumparts
where we have used the notation $\vec e(\vec k)=\vec \epsilon_+(\vec k)=\vec \epsilon_-(\vec k)^\star$ and the creation-anihilation operators are linked through $\hat  a(\vec k)=i\hat  C_{\vec g_{\vec k,+}}$ and $\hat  b^\dag(\vec k)=i\hat  C^\dag_{\vec g_{\vec k,+}}$. The expression \eref{F_RS BB} coincides with \cite[equation (7)]{BialynickiBirula1998} which confirms that the quantization we have defined in \eref{BBquantization1}-\eref{BBquantization2} coincides with the one used by BB e.g. in \cite{BialynickiBirula1996,BialynickiBirula1998,BialynickiBirula2013,BialynickiBirula2012,BialynickiBirula2012a,BialynickiBirula2017,BialynickiBirula2006}.

The representation of the energy observable and the generator of the time evolution will be discussed in \Sref{Dynamics}.

\section{Equivalence between the LP and the BB quantizations --- isomorphism}
\label{iso}
In \Sref{eq LP-m}, we have shown that the Coulomb gauge quantizations in position and momentum representations are equivalent and linked through explicit relations given by the classical isomorphism $\mathcal{M}$. In the same spirit, we will now introduce a classical isomorphism $\mathcal{I}$ between the LP and the BB Hilbert spaces in order to show that the two constructed quantized theories are equivalent. We will also show how one can pass from one theory to the other through explicit transformations involving the classical isomorphism.

We start by defining the map
\begin{eqnarray}
\mathcal{I}:\mathcal{H}&_{LP}&\rightarrow\mathcal{H}_{BB}\nonumber\\
&\vec\psi&\mapsto\vec F_{BB}=\mathcal{I}\vec\psi:=i\sqrt{\hbar}\Omega^{1/2}\vec\psi,\label{iso I}
\end{eqnarray}
which is extended to the Fock space by also defining $\mathcal{I}|\varnothing\rangle_{LP}=|\varnothing\rangle_{BB}$. The inverse transformation is given by
\begin{eqnarray}
\mathcal{I}^{-1}:\mathcal{H}&_{BB}&\rightarrow\mathcal{H}_{LP}\nonumber\\
&\vec F_{BB}&\mapsto\vec\psi=\mathcal{I}^{-1}\vec F_{BB}:=-\frac{i}{\sqrt{\hbar}}\Omega^{-1/2}\vec F_{BB}.
\end{eqnarray}
One can check that $\mathcal{I}$ indeed links the two representations by computing directly
\numparts
\begin{eqnarray}
i\sqrt{\hbar}\Omega^{1/2}\vec\psi&=\frac{i}{\sqrt{2}}\Omega^{1/2}\left( \sqrt{\varepsilon_0}\Omega^{1/2} \vec A -i\sqrt{\varepsilon_0}\Omega^{-1/2}\vec E \right)\\
&=i\sqrt{\frac{\varepsilon_0}{2}}\left( \Omega\vec A -i\vec E \right)\\
&=\sqrt{\frac{\varepsilon_0}{2}}\left(i c\Lambda\nabla\times\vec A+\vec E\right)\\
&=\frac{1}{\sqrt{2}}\left(\sqrt{\varepsilon_0}\vec E +\frac{i}{\sqrt{\mu_0}}\Lambda\vec B\right)\\
&=\frac{1}{\sqrt{2}}\left(\vec{\mathcal{Q}}_{BB} +i \vec{\mathcal{P}}_{BB}\right)=\vec F_{BB},
\end{eqnarray}
\endnumparts
where we have used the relation $\Omega=c\Lambda\nabla\times$ and $\Lambda^{-1}=\Lambda$.
The equivalence of their respective scalar products follows immedialtely from the definitions
\numparts
\begin{eqnarray}
\langle\mathcal{I}\vec\psi|\mathcal{I}\vec\psi'\rangle_{BB}&=\frac{1}{\hbar}\int_{\mathbb{R}^3}d^3x\ (\mathcal{I}\vec\psi)^\star\cdot\Omega^{-1}\mathcal{I}\vec\psi'\label{scalar prod eq 1}\\
&=\int_{\mathbb{R}^3}d^3x\ \Omega^{1/2}\vec\psi^\star\cdot\Omega^{-1}\Omega^{1/2}\vec\psi'\label{scalar prod eq 2}\\
&=\int_{\mathbb{R}^3}d^3x\ \vec\psi^\star\cdot\vec\psi'\label{scalar prod eq 3}=\langle\vec\psi|\vec\psi'\rangle_{LP}\label{scalar prod eq 4},
\end{eqnarray}
\endnumparts
where we have used the selfadjoint property of $\Omega^{1/2}$ to obtain \eref{scalar prod eq 3}.

We can express the link between the creation-anihilation operators of both representations by starting e.g. with a single-photon state in the LP representation
\begin{eqnarray}
\hat B_{\vec\psi}^\dag|\varnothing\rangle_{LP}=|\vec\psi\rangle_{LP},
\end{eqnarray}
which can be rewritten using the isomorphism $\mathcal{I}$ as
\numparts
\begin{eqnarray}
|\vec \psi\rangle_{LP}&=|\mathcal{I}^{-1}\vec F_{BB}\rangle_{LP}\\
&=\mathcal{I}^{-1}|\vec F_{BB}\rangle_{BB}\\
&=\mathcal{I}^{-1}\hat C_{\vec F_{BB}}^\dag|\varnothing\rangle_{BB}\\
&=\mathcal{I}^{-1}\hat C_{\vec F_{BB}}^\dag \mathcal{I}|\varnothing\rangle_{LP}.
\end{eqnarray}
\endnumparts
From this one can identify
\begin{eqnarray}
\hat B^\dag_{\vec\psi}=\mathcal{I}^{-1}\hat C^\dag_{\vec F_{BB}}\mathcal{I}=\mathcal{I}^{-1}\hat C^\dag_{\mathcal{I}\vec \psi}\mathcal{I},
\end{eqnarray}
and therefore the other relations are 
\begin{eqnarray}
\hat C^\dag_{\vec F_{BB}}=\mathcal{I}\hat B^\dag_{\vec\psi}\mathcal{I}^{-1},\:\:\:\:\:\:\:\:\:
\hat B_{\vec\psi}=\mathcal{I}^{-1}\hat C_{\vec F_{BB}}\mathcal{I}, \:\:\:\:\:\:\:\:\:
\hat C_{\vec F_{BB}}=\mathcal{I}\hat B_{\vec\psi}\mathcal{I}^{-1}.
\end{eqnarray}
The equivalence of electromagnetic quantities can be verified for instance for the electric field 
\numparts
\begin{eqnarray}
\mathcal{I}\vec{\hat E}_{LP}\mathcal{I}^{-1}&=i\sqrt{\frac{\hbar}{2\varepsilon_0}}\int_{\mathbb{R}^3}d^3k\sum_{\sigma=\pm}\omega_{\vec k}^{1/2}\left(\vec\phi_{\vec k,\sigma}(\vec x)\mathcal{I}\hat  B_{\vec\phi_{\vec k,\sigma}}\mathcal{I}^{-1}-\vec\phi^\star_{\vec k,\sigma}(\vec x)\mathcal{I}\hat  B^\dag_{\vec\phi_{\vec k,\sigma}}\mathcal{I}^{-1}\right)\nonumber\\\\
&=i\sqrt{\frac{\hbar}{2\varepsilon_0}}\int_{\mathbb{R}^3}d^3k\sum_{\sigma=\pm}\omega_{\vec k}^{1/2}\left(\vec\phi_{\vec k,\sigma}(\vec x)\hat  C_{\mathcal{I}\vec\phi_{\vec k,\sigma}}-\vec\phi^\star_{\vec k,\sigma}(\vec x)\hat  C^\dag_{\mathcal{I}\vec\phi_{\vec k,\sigma}}\right)\\
&=\frac{1}{\sqrt{2\varepsilon_0}}\int_{\mathbb{R}^3}d^3k\sum_{\sigma=\pm}\left(\vec g_{\vec k,\sigma}(\vec x)\hat  C_{\vec g_{\vec k,\sigma}}+\vec g^\star_{\vec k,\sigma}(\vec x)\hat  C^\dag_{\vec g_{\vec k,\sigma}}\right)\\
&=\frac{1}{\sqrt{2\varepsilon_0}}\left(\vec{\hat F}_{BB}(\vec x)+\vec{\hat F}^\dag_{BB}(\vec x)\right)\equiv\vec{\hat E}_{BB}(\vec x),
\end{eqnarray}
\endnumparts
where we have used the basis $\vec g_{\vec k,\sigma}=i\sqrt{\hbar\omega_{\vec k}}\vec\phi_{\vec k,\sigma}$ of $\mathcal{H}_{BB}$ and we have identified $\vec{\hat F}_{BB}$ and $\vec{\hat F}_{BB}^\dag$ by their definitions \eref{BB field op1},\eref{BB field op2}. This expression indeed coincides with \eref{E field BB} for the electric field.

However, one has to be careful when transforming other objects like the field operators $\vec{\hat\Psi}$ and $\vec{\hat F}_{BB}$. Indeed, $\vec{\hat F}_{BB}\neq\mathcal{I}\vec{\hat\Psi}\mathcal{I}^{-1}$ since if one decomposes the LP field operator in terms of creation-anihilation operators, it yields
\numparts
\begin{eqnarray}
\mathcal{I}\vec{\hat\Psi}(\vec x)\mathcal{I}^{-1}&=\int_{\mathbb{R}^3}d^3k\sum_{\sigma=\pm}\vec\phi_{\vec k,\sigma}(\vec x)\mathcal{I}\hat  B_{\vec\phi_{\vec k,\sigma}}\mathcal{I}^{-1}\\
&=\int_{\mathbb{R}^3}d^3k\sum_{\sigma=\pm}\vec\phi_{\vec k,\sigma}(\vec x)\hat  C_{\vec g_{\vec k,\sigma}}\\
&=-i\int_{\mathbb{R}^3}d^3k\sum_{\sigma=\pm}(\hbar\omega_{\vec k})^{-1/2}\vec g_{\vec k,\sigma}(\vec x)\hat  C_{\vec g_{\vec k,\sigma}}\\
&=\frac{-i}{\sqrt{\hbar}}\Omega^{-1/2}\vec{\hat F}_{BB}.\label{iso field op}
\end{eqnarray}
\endnumparts

\section{Dynamics and BB quantum Hamiltonian}
\label{Dynamics}

The Hamiltonian in a quantum theory plays two roles: it is the operator associated to the total energy observable of the system and it is the generator of the dynamics of the states in the Schrödinger representation. The construction of the dynamics within the Fock space formulation is done as follows: one starts with the one-quantum subspace, which coincides with the classical Hilbert space $\mathcal{H}$, and whose time evolution is given by the classical one. From this, a one-quantum operator is identified as the generator of the dynamics. For Maxwell's equations, the generator of the dynamics is $\hbar\Omega$ for all representations we discussed before ($\hbar$ has been added here in relation to \eref{max-LP}, \eref{time ev FBB} or \eref{time ev calFBB} in order to match the dimension required by a Schrödinger equation in the quantized model). The next step is to extend this operator $\hbar\Omega$, which is defined in the single-photon subspace, to the entire Fock space $\mathbb{F}^{\mathfrak{B}}(\mathcal{H})$.   This  can be done in general with the map denoted $\rmd\Gamma(\hbar\Omega)$ \cite{Honegger2014} and defined by its action on the monomials $| \eta_1\otimes_S\eta_2\otimes_S \ldots \otimes_S  \eta_n  \rangle $:
\begin{eqnarray} 
\rmd\Gamma(\hbar\Omega) ~| \eta_1\otimes_S\eta_2\otimes_S \ldots \otimes_S  \eta_n  \rangle 
&:=|\hbar\Omega \eta_1\otimes_S\eta_2\otimes_S \ldots \otimes_S  \eta_n  \rangle \nonumber\\
& \:\: + | \eta_1\otimes_S\hbar\Omega \eta_2\otimes_S \ldots \otimes_S  \eta_n  \rangle  \nonumber\\
& \:\: +\dots+\nonumber \\ 
& \:\:+ | \eta_1\otimes_S\ \eta_2\otimes_S \ldots \otimes_S  \hbar\Omega\eta_n  \rangle.
\end{eqnarray}
If $\phi_\kappa$ is an orthonormal basis, with respect to $\langle\cdot|\cdot\rangle_{\mathcal{H}}$, of eigenfunctions of   $\Omega \phi_\kappa  = \omega_\kappa  \phi_\kappa$, the map $\rmd\Gamma(\hbar\Omega)$ can be expressed as \cite[p. 440]{Honegger2014}
\begin{equation} 
\rmd\Gamma(\hbar\Omega) =\sum_\kappa \hbar \omega_\kappa \hat B^\dag_{\phi_\kappa} \hat B_{\phi_\kappa}.
\end{equation}
We apply this procedure to the different Hilbert spaces defined in the preceding sections starting with the LP representation 
\numparts
\begin{eqnarray}
\rmd\Gamma_{LP}(\hbar\Omega)&=\int_{\mathbb{R}^3}d^3k \sum_{\sigma=\pm}\hbar\omega_{\vec k}\hat B^\dag_{\vec\phi_{\vec k,\sigma}}\hat B_{\vec\phi_{\vec k,\sigma}}.
\end{eqnarray}
\endnumparts
We see that it is equal to the quantized total energy operator which can be written as
\begin{eqnarray}
\hat H_{LP}&=\hbar\int_{\mathbb{R}^3} d^3x\ \vec{\hat\Psi}^{\dag}\cdot\Omega\vec{\hat\Psi}=\int_{\mathbb{R}^3}d^3k\sum_{\sigma=\pm}\hbar\omega_{\vec k}\hat B^\dag_{\vec\phi_{\vec k,\sigma}}\hat B_{\vec\phi_{\vec k,\sigma}},
\end{eqnarray}
showing that the total energy is the generator of the dynamics in the Coulomb gauge formulation of the quantized electromagnetic field. 

One can apply the same procedure to the BB formulation using the basis $\{\vec g_{\vec k,\sigma}\}$ and one obtains
\numparts
\begin{eqnarray}
\rmd\Gamma_{BB}(\hbar\Omega)&=\int_{\mathbb{R}^3}d^3k \sum_{\sigma=\pm}\hbar\omega_{\vec k}\hat C^\dag_{\vec g_{\vec k,\sigma}}\hat C_{\vec g_{\vec k,\sigma}},
\end{eqnarray}
\endnumparts
which coincides with the total energy
\numparts
\begin{eqnarray}
\hat \mathcal{E}_{tot}&=\int_{\mathbb{R}^3} d^3x\ \vec{\hat F}_{BB}^{\dag}\cdot\vec{\hat F}_{BB}\\
&=\sum_{\kappa,\kappa'}\left(\int_{\mathbb{R}^3} d^3x\ \vec g^\star_\kappa(\vec x)\cdot\vec g_{\kappa'}(\vec x)\right)\hat C^\dag_{\vec g_\kappa}\hat C_{\vec g_{\kappa'}}\\
&=\hbar\sum_{\kappa,\kappa'}\langle\vec g_{\kappa}|\Omega\vec g_{\kappa'}\rangle_{BB}\ \hat C^\dag_{\vec g_{\kappa}}\hat C_{\vec g_{\kappa'}}\\
&=\sum_\kappa\hbar\omega_\kappa\ \hat C^\dag_{\vec g_{\kappa}}\hat C_{\vec g_{\kappa}}\equiv\hat H_{BB},\label{H creation anihil}
\end{eqnarray}
\endnumparts
where we have introduced the more compact notations $\kappa\equiv(\vec k,\sigma)$ and $\sum_\kappa\equiv\int_{\mathbb{R}^3}d^3k\sum_{\sigma=\pm}$. This confirms that in the quantized theory, the generator of the dynamics is the quantized operator of the total energy in the BB representation and not the quantization of the classical Hamilton function $K_{BB}$ \eref{KBB}. The time evolution of the states is given by the unitary operator $\hat U_{BB}(t)=e^{-i\hat H_{BB}t}$, and it leads e.g. when applied to single-photon states to the expressions of the form \cite[equation (4.20)]{BialynickiBirula1996}. One can further check that the Hamiltonians written in their respective representations can be recovered from one to the other using the isomorphism (see \eref{iso field op})
\numparts
\begin{eqnarray}
\mathcal{I}\hat H_{LP}\mathcal{I}^{-1}&=\hbar\int_{\mathbb{R}^3} d^3x\ \mathcal{I}\vec{\hat\Psi}^{\dag}\mathcal{I}^{-1}\cdot\Omega\mathcal{I}\vec{\hat\Psi}\mathcal{I}^{-1}\\
&=\int_{\mathbb{R}^3} d^3x\ \Omega^{-1/2}\vec{\hat F}_{BB}^{\dag}\cdot\Omega^{1/2}\vec{\hat F}_{BB}\\
&=\int_{\mathbb{R}^3} d^3x\ \vec{\hat F}_{BB}^{\dag}\cdot\vec{\hat F}_{BB}=\hat H_{BB}.
\end{eqnarray}
\endnumparts
The equivalence of the generators of the dynamics guarantees that the isomorphism is preserved during the time evolution and that the two theories are completely equivalent.

\section{Concluding remarks}

\begin{itemize}
\item In the literature, the relation between the LP field and the RS vector has been discussed for instance in \cite[section 5.3]{BialynickiBirula1996} where it was stated that the two objects can be linked but only from BB to LP i.e. for every $\vec F_{BB}\in\mathcal{H}_{BB}$, there should exist a well defined $\vec\psi\in\mathcal{H}_{LP}$ but a doubt was expressed about the reverse. The singularity of the inverse transformation and the nonintegrability of a kernel of the form $|\vec x -\vec x'|^{-7/2}$ could prevent a bijection between the two representations. Our present formulation of the isomorphism shows that there is no such difficulty. Our argument is based on the fact that $\Omega$ is a positive selfadjoint operator and both $\Omega^{\pm1/2}$ are well defined and selfadjoint in the Hilbert space. Concerning the representation as integral operators, it was shown e.g. in \cite{Stein1971,Landkof2011,Kwasnicki2017,Lischke2020} that they can be written for any field $\vec v(\vec x)$ as
\begin{equation}
\left[\Omega^{-1/2}\vec v\right](\vec x)=\frac{\pi}{\sqrt{c}}\int_{\mathbb{R}^3}d^3x' \frac{\vec v(\vec x')}{\left( 2\pi |\vec x -\vec x'|\right)^{5/2}},
\end{equation}
and 
\begin{equation}
\left[\Omega^{1/2}\vec v\right](\vec x)=\frac{3\sqrt{2c}}{16\pi^{3/2}}\mathcal{P}\int_{\mathbb{R}^3}d^3x' \frac{\vec v(\vec x)-\vec v(\vec x')}{ |\vec x -\vec x'|^{7/2}},
\end{equation}
where $\mathcal{P}$ stands for the Cauchy principal value. Consequently, the isomorphism guarantees a one-to-one correspondence between the states from one Hilbert space to the other.
\item In a recent article \cite{BialynickiBirula2020}, the concept of fidelity for photons was discussed using the BB quantization in both position and momentum space representations. The BB momentum representation can be written from the BB position representation as follows
\numparts
\begin{eqnarray}
\tilde f_m(\vec k,\sigma)&:=\sqrt{\frac{(2\pi)^3}{\hbar c}}\int_{\mathbb{R}^3}d^3x\ \vec\phi_{\vec k,\sigma}^\star(\vec x)\cdot\vec F_{BB}(\vec x),\label{BBm iso1}\\
\vec F_{BB}(\vec x)&:=\sqrt{\frac{\hbar c}{(2\pi)^3}}\int_{\mathbb{R}^3}d^3k\sum_{\sigma=\pm} \tilde f_m(\vec k,\sigma)\vec\phi_{\vec k,\sigma}(\vec x)\label{BBm iso2},
\end{eqnarray}
\endnumparts
and it defines a BB momentum Hilbert space
\begin{equation}
\mathcal{H}_{BBm}:=\{ \tilde f_m(\vec k,\sigma) | \langle\tilde f_m|\tilde f_m\rangle_{BBm}<\infty\},\label{BBm H space}
\end{equation}
with the BB momentum weighted scalar product \cite[equation (2)]{BialynickiBirula2020}
\begin{equation}
\langle\tilde f_{m1}|\tilde f_{m2}\rangle_{BBm}:=\int_{\mathbb{R}^3}\frac{d^3k}{(2\pi)^3|\vec k|}\sum_{\sigma=\pm}\tilde f_{m1}^\star(\vec k,\sigma)\tilde f_{m2}(\vec k,\sigma).\label{BBm scalar product}
\end{equation}
One can check that equations \eref{BBm iso1},\eref{BBm iso2} and \eref{BBm H space} provide a Hilbert space isomorphism between $\mathcal{H}_{BB}$ and $\mathcal{H}_{BBm}$. Consequently, if one defines the fidelity in momentum space to be \cite[equation (18)]{BialynickiBirula2020})
\begin{equation}
\mathfrak{F}_m:=\frac{|\langle\tilde f_{m1}|\tilde f_{m2}\rangle_{BBm}|^2}{\langle\tilde f_{m1}|\tilde f_{m1}\rangle_{BBm}\langle\tilde f_{m2}|\tilde f_{m2}\rangle_{BBm}},\label{fid mom}
\end{equation}
the isomorphism implies immediately that 
\begin{eqnarray}
\mathfrak{F}_m=\frac{|\langle\vec F_{BB1}|\vec F_{BB2}\rangle_{BB}|^2}{\langle\vec F_{BB1}|\vec F_{BB1}\rangle_{BB}\langle\vec F_{BB2}|\vec F_{BB2}\rangle_{BB}}=:\mathfrak{F}_{BB},\label{fid pos}
\end{eqnarray}
where $\mathfrak{F}_{BB}$ is the fidelity in position space. In \cite{BialynickiBirula2020}, a comparison is made with another conceivable definition of fidelity
\begin{eqnarray}
\mathfrak{F}'_{BB}:=\frac{|\langle\vec F_{BB1}|\vec F_{BB2}\rangle|^2}{\langle\vec F_{BB1}|\vec F_{BB1}\rangle\langle\vec F_{BB2}|\vec F_{BB2}\rangle},\label{wrong fid}
\end{eqnarray}
with a scalar product $\langle\cdot|\cdot\rangle$ without the weight $\Omega^{-1}$
\begin{equation}
\langle\vec F_{BB1}|\vec F_{BB2}\rangle:=\int_{\mathbb{R}^3}d^3x\ \vec F_{BB1}^\star(\vec x)\cdot\vec F_{BB2}(\vec x)\neq\langle\vec F_{BB1}|\vec F_{BB2}\rangle_{BB}.
\end{equation}
The results discussed in the present work indicate that $\mathfrak{F}_m$ and $\mathfrak{F}_{BB}$ should be taken as the measure of fidelity and not $\mathfrak{F}'_{BB}$. The main point is that the scalar product of the Hilbert space $\mathcal{H}_{BB}$ should be used if one works in the BB representation. Furthemore, through the isomorphism $\mathcal{I}:\mathcal{H}_{LP}\rightarrow\mathcal{H}_{BB}$ the fidelities $\mathfrak{F}_m$ and $\mathfrak{F}_{BB}$ coincide with the one defined in the LP as well as in the standard momentum representation.

If one uses the Hilbert space $\mathcal{H}_{BB}$ that we defined in \eref{H BB}, the fidelity for photons is the same whether it is measured in position or momentum space as expected for a physical property which should always be independent of the chosen representation for the states.

\item The isomorphism between the LP and the BB representations can also be interpreted in a reversed way. If we start with the LP representation it has two weak features: it relays on the choice of a particular gauge and it is not manifestly Lorentz invariant. The isomorphism $\mathcal{I}$ provides a simple way to modify the formulation to obtain a gauge independent and manifestly Lorentz invariant representation.

\item The relation between $\vec F_{BB}$ and  $\vec F_{RS}$ (and thus also with $\vec E$ and $\vec B$) is non-local, since it involves the projection into the positive and negative helicity parts, which is a non-local operation. The non-local character, which was considered a weak feature of the LP representation, is thus also present in the BB representation. One might surmise that it is likely a general feature of the quantum field theories for photons.

\item The field $\Omega^{1/2} \vec \psi$ had been already introduced in the book  by L. Mandel and E. Wolf \cite[equation (12.11-32) ]{Mandel1995}, and called the ``energy wave function'', as well as in \cite[equation (2.15)]{Sipe1995}. This denomination has been also adopted for the six component vector $\Psi_{BB}$ in \cite[eq (10.8)]{Keller2005},\cite[p.3]{Smith2007}.
These denominations are however somewhat misleading. As it was shown in   \cite[equation (12)]{BialynickiBirula1998}, the mean value of the local energy density can be expressed as
\numparts
\begin{eqnarray} 
\langle\hat{\mathcal{E}}_{elm}\rangle&= |\vec F_{RS}^{(h+)}|^2  +|\vec F_{RS}^{(h-)}|^2  = |\vec F_{BB}^{(h+)}|^2  +|\vec F_{BB}^{(h-)}|^2 \\
& = |\Psi_{BB}|^2.
\end{eqnarray}
\endnumparts
However, $|\Psi_{BB}|^2$ is not a Born rule probability density for the local energy observable. It appears naturally in the mean value of the local energy, but it is not its probability. 
 Since the concept of wave function carries most often the connotation of a Born rule probability,
we prefer to use the denomination {\it Bia{\l}ynicki-Birula bispinor} for $\Psi_{BB}$ and {\it Bia{\l}ynicki-Birula vector } for $\vec F_{BB}$.

\end{itemize}

\section{Conclusion}

In this article, we have explored different approaches of the quantization of the free electromagnetic field in position space that use the LP field and the BB vector. The key result was to show that they are linked by an isomorphism of their respective Hilbert spaces that we explicitely wrote using the frequency and the helicity operators. With this relation we were able to show that the quantum theories built from them through a Fock space quantization are equivalent, meaning that they predict the same physical results. This equivalence is preserved during the time evolution. In the literature, both the LP field and BB vectors were used for different practical problems and using the results of the present work, any calculation made from one representation can be easily translated into the other using the isomorphism $\mathcal{I}$.

\ack
This work was supported by the ``Investissements d'Avenir'' program, project ISITE-BFC / IQUINS (ANR-15-IDEX-03), QUACO-PRC (ANR-17-CE40-0007-01) and the EUR-EIPHI Graduate School (17-EURE-0002). We also acknowledge support from the European Union's Horizon 2020 research and innovation program under the Marie Sklodowska-Curie grant agreement No. 765075 (LIMQUET).

\section*{References}

\bibliographystyle{unsrt}
\bibliography{JPhysA-118372_revised}
\end{document}